\documentstyle[12pt]{article}
\newcommand {\pl}{\partial}
\newcommand {\p} {\phi}
\newcommand {\vp}{\varphi}

\newcommand {\al}{\alpha}
\newcommand {\be}{\beta}
\newcommand {\ga}{\gamma}
\newcommand {\Ga}{\Gamma}

\newcommand {\la}{\lambda}
\newcommand {\La}{\Lambda}

\newcommand {\del}  {\delta}
\newcommand {\Del}  {\Delta}

\newcommand {\half}{ {\frac{1}{2}} }

\newcommand {\sqg} {\sqrt{g}}

\newcommand {\Lcal}{{\cal L}}

\newcommand {\Dcal}{{\cal D}}

\newcommand {\Zhat}{{\hat Z}}
\newcommand {\Gahat}{{\hat \Gamma}}

\newcommand {\PL}  {{\hbar}}
\newcommand {\PLinv}  {{\frac{1}{\hbar}}}
\newcommand {\intx} {{\int d^2x}}

\newcommand {\ra} {\rightarrow}
\newcommand {\pr}   {{\quad .}}
\newcommand {\com}  {{\quad ,}}
\newcommand {\q}    {\quad}
\newcommand {\qq}   {\quad\quad}
\newcommand {\qqq}   {\quad\quad\quad}
\newcommand {\qqqq}   {\quad\quad\quad\quad}

\newcommand {\nn}    {\nonumber}
\newcommand {\vs}[1]  { \vspace*{#1 cm} }

\newcounter{eq}
\newcounter{sc}

\newcommand {\MPL}  {Mod.Phys.Lett.}
\newcommand {\NP}   {Nucl.Phys.}
\newcommand {\PR}   {Phys.Rev.}

\newcommand {\PTP}  {Prog.Theor.Phys.}

\begin{document}
\title{   Classical Solution of Two Dimensional $R^2$-Gravity
           and Cross-Over Phenomenon
          \thanks{US-94-03}
                                 }
\author{
          S. ICHINOSE
          \thanks{ E-mail address:\ ichinose@u-shizuoka-ken.ac.jp}\\
          Department of Physics, Universuty of Shizuoka,      \\
          Yada 52-1, Shizuoka 422, Japan                             \\
          N. TSUDA
          \thanks{ E-mail address:\ ntsuda@theory.kek.jp }\\
          Department of Physics,Tokyo Institute of Technology, \\
          Oh-okayama,Meguro-ku,Tokyo 152,  Japan \\
          and T. YUKAWA
          \thanks{ E-mail address:\ yukawa@theory.kek.jp }              \\
          National Laboratory for High Energy Physics(KEK), \\
          Tsukuba,Ibaraki 305 ,Japan                                  \\
                          }
\date{  February, 1995 }
\maketitle
\setlength{\baselineskip}{0.54cm}
\begin{abstract}
Two dimensional quantum R$^2$-gravity and its phase structure are
examined in the semiclassical approach and
compared with the results of the numerical simulation.
Three phases are succinctly characterized by the effective action.
A classical solution
of R$^2$-Liouville equation is obtained by use of the solution of
the ordinary Liouville equation.
The partition function is obtained analytically.
A toatal derivative term (surface term) plays
an important role there.
It is shown that the classical solution  can sufficiently
account for the cross-over
transition of the surface property seen in the numerical simulation.
\end{abstract}
\section{Introduction}
Importance of the semiclassical approach to the  quantum
gravity has long been known. (For a recent review, see
\cite{BOS}.)
It is true as well in
the two dimensional (2d) quantum gravity.
Liouville theory, which is
equivalent to the 2d quantum gravity in the conformal
gauge, has be treateded  semiclassically \cite{DJ,SEI}.
In this paper we study 2d quantum $R^2$-gravity in the similar manner.
The motivations for studying this model can be said as follows. Firstly,
the ordinary 2d gravity is essentially based on the lagrangian:\
$\Lcal=\sqg (\frac{1}{2\ga}R\frac{1}{\Del}R+\frac{1}{G}R+\mu)$. Because
Einstein term,$\intx\sqg R$\ ,is a topological quantity, the dynamical effect
comes only from the induced part ,$R\frac{1}{\Del}R$\ . The lowest
derivative-order 'kinetic' term
,made purely of metric, is $R^2$. If higher-derivative
terms have some meaning in 2d quantum gravity, this model is worthy of study
as the simplest one. Secondly, the simulation data
of $R^2$-gravity, with high statistics,
has recently appeared. This theory is a good testing model of
the quantum gravity that can be compared with the numerical experiment.
We can examine how
some important procedures, such as (infra-red and ultra-violet)
regularization and renormalization, of the field theory work in the model.

$R^2$-gravity,for Lorentzian metric,
was first quantumly treated by T.Yoneya\cite{Y},\linebreak
where
Hamilton-Jacobi equation in the superspace approach is exactly solved.
Its importance as an regularization (of the ultra-violet behaviour)
was suggested
in \cite{KPZ}. One of us (S.I.) has shown its renormalizability using the
background-field method\cite{SI} and obtained some renormalization-group
beta-functions. Kawai and Nakayama(KN)\cite{KN} have treated the system
based on the conformal field theory. Their approach will be compared
with the present one in sect.5.

Another interesting approach to the quantum gravity is the lattice simulation.
Since the method of the dynamical triangulation was invented for the Euclidean
quantum gravity \cite{ADF,D,KKM}, the non-perturbative aspect of
the quantum gravity has been vigorously analysed these 10 years.
By this approach, the effect of $R^2$-term
was examined by \cite{BK} in the early stage of the development
of the simulation.
Recently a cross-over phenomenon
of the surface from the fractal phase to the 'flat' phase was clearly observed
\cite{TY,YTS}.
The computer simulation of quantum gravity
has been now greatly improved.
Especially  data of 2d quantum gravity
become so accurate that they can be closely compared
with the analytical prediction.
We examine  the recent computer-simulation data of the 2d R$^2$-gravity
and present its theoretical interpretation, especially focus our attention on
the cross-over transition.

The semiclassical approach  was intensively applied
to the quantization around an extended
object (soliton, kink,instanton ,etc.) \cite{RA}.
The advantage of this approach
is that the whole physical situation is simply viewed in an effective
action. In this approach
the central role is played by the classical solution.
Non-perturbative effects are taken into account
by incorporating  the non-trivial classical vacuum (Liouville solution in
the present case), while the fluctuation around
the solution is treated perturbatively. In analyzing the 2d $R^2$-quantum
gravity semi-classically, we must first find the appropriate
classical solution.

We take the Euclidean action,
\begin{eqnarray}
& S_{tot}=S_{gra}+S_m\com \q
 S_{gra}[g;G,\be,\mu]=\intx\sqg (\frac{1}{G} R-\be R^2-\mu)\com
                                                              & \nn\\
& S_m[g,\Phi;c_m]=-\intx\sqg (\half\sum_{i=1}^{c_m}\pl_a\Phi_i\cdot
g^{ab}\cdot \pl_b\Phi_i)\com\q (\ a,b=1,2\ )\com            & \label{1.1}
\end{eqnarray}
under the fixed area condition
$ A=\intx \sqg\ $. Here
$G$\ is the gravitaional coupling constant, $\mu$\ is the cosmological
constant ,
$\be$\ is the coupling strength for $R^2$-term and $\Phi$\ is the $c_m$-
components scalar matter fields.

\section{Semiclassical Quantization}
By taking the conformal-flat gauge,\
$g_{ab}=\ e^{\vp}\ \del_{ab}$\ ,
the action~(\ref{1.1})
gives us,after integrating out the matter fields and Faddeev-Popov ghost,
the following partition function\cite{P}.
\begin{eqnarray}
& \int\frac{\Dcal g\Dcal\Phi}{V_{GC}}\{exp\PLinv S_{tot}\}~\del(\intx\sqg-A)
=exp\PLinv (\frac{8\pi(1-h)}{G}-\mu A)\times Z[A]\com & \nn\\
& Z[A]\equiv\int\Dcal\vp~ e^{+\frac{1}{\PL}
S_0[\vp]}~\del(\intx ~e^\vp - A)\com &   \label{3.3}\\
& S_0[\vp]=\intx\ (\frac{1}{2\ga}\vp\pl^2\vp
-\be~e^{-\vp}(\pl^2\vp)^2 +\frac{\xi}{2\ga}\pl_a(\vp\pl_a\vp)\ )\com
\q \frac{1}{\ga}=\frac{1}{48\pi}(26-c_m)\com & \label{3.2}
\end{eqnarray}
where  the  relations for
Einstein term and the cosmological term:\
$\intx\sqg R =8\pi (1-h),\ h=\mbox{number of handles},
\intx\sqg =A $\ ,are used.
\footnote{
The sign for the action is different from the usual convention as seen in
(\ref{3.3}).
}
$V_{GC}$\ is the gauge volume due to the general coordinate invariance.
$\xi$\ is a free parameter. The total derivative term generally appears when
integrating out the anomaly equation
\ $\del S_{ind}[\vp]/\del\vp=\frac{1}{\ga}\pl^2\vp\ $.
This term turns out to be very important.
\footnote{
The uniqueness of this term, among all possible total derivatives, is shown
in Discussions(sect.6).
}
We consider the manifold of a fixed topology of
the sphere ,$h=0$, and with the finite area $A$.
Furthermore we consider the case $\ga>0\ (c_m<26)$.
\footnote{
This is for the comparison with the 'classical limit' $c_m\ra -\infty$.
We can do the same analysis for $\ga<0$\ without any difficulty.
}
\ $\PL$\ is  Planck constant.
\footnote{
In this section only,we explicitly write $\PL$\ (Planck constant) in order
to show the perturbation structure clearly.
}

\qq
Let us describe the thermo-dynamical consideration which will be crucial
in later discussions.
The Laplace transform of (\ref{3.3}) is written as
\begin{eqnarray}
{\Zhat}[\la]=
\int_0^{\infty} Z[A]e^{-\la A/\PL}~dA
\ =\int\Dcal\vp~exp[~
+\frac{1}{\PL}\{ S_o[\vp]-\la\ \intx ~e^\vp \} ]\pr
                                                      \label{3.4a}
\end{eqnarray}
$Z[A]$\ is  the micro-canonical partition function with
the area $A$\ ,
while $\Zhat[\la]$\ is the grand-canonical partiton function
with the chemical potential $\la$\ . In the grand-canonical case, the average
area is controled by fixing $\la$\ through the relation,
\begin{eqnarray}
<A_{op}>=\frac{1}{\Zhat}\frac{d}{d(-\la/\PL )}\Zhat[\la]
\equiv <\intx e^{\vp}>_{\Zhat}\com\q
A_{op}\equiv\intx~e^\vp\pr                  \label{3.4c}
\end{eqnarray}

\qq Conversely, the micro-canonical partiton function can be obtained from
$\Zhat[\la]$\ by the inverse Laplace transformation,
\begin{eqnarray}
Z[A]=\int\frac{d\la}{\hbar}\Zhat[\la]~e^{+\la A/\hbar}\pr  \label{3.4cx}
\end{eqnarray}
The integral should be carried out along an appropriate contour parallel
to the imaginary axis. We write  $\Zhat[\la]$ as
\begin{eqnarray}
&S_\la[\vp]\equiv S_0[\vp]-\la\intx~e^\vp\ &\nn\\
&=\intx\ (\frac{1}{2\ga}\vp\pl^2\vp -\be~e^{-\vp}(\pl^2\vp)^2\
+\frac{\xi}{2\ga}\pl_a(\vp\pl_a\vp)\
 -\la~e^\vp\ )\com    &\label{3.4g}\\
&\Zhat[\la]=\int\Dcal\vp~exp~\{\PLinv S_\la[\vp]\}
\equiv\ exp~\PLinv\Gahat[\la]\com &   \nn
\end{eqnarray}
where $\Gahat(\la)$\ is the effective action induced by $S_\la[\vp]$.
It can be calculated loop-wise
by the semiclassical expansion\ :\
$\vp(x)=\vp_c(x;\la)+\sqrt{\PL}~\psi(x)$\ ,
with taking the solution
of the classical field equation :\
$\left.\frac{\del}{\del\vp}S_\la[\vp]\right|_{\vp_c}=\ 0$\ ,
as the background field.
Then  we have
\begin{eqnarray}
& \Zhat[\la]=~
exp~\PLinv S_\la[\vp_c]\times \int \Dcal\psi~
exp\{\frac{1}{2}\frac{\del^2S_\la}{\del\vp^2}|_{\vp_c}\psi\psi
                   +O(\sqrt{\PL}) \}    &         \nn\\
& \equiv exp\{\PLinv\Gahat^0[\la]+\Gahat^1[\la]+
O(\PL)\}\com                                   &  \label{3.4j}    \\
& \Gahat[\la]=\Gahat^0[\la]+\PL\Gahat^1[\la]
+O(\PL^2 )\com\q
\Gahat^0[\la]\equiv S_\la[\vp_c]\com    &     \nn
\end{eqnarray}
where $\Gahat^n[\la],(n\geq 1),$\ is the quantum effects contributed
from n-loop diagrams.

\qq Writing the integrand of (\ref{3.4cx}) as
\begin{eqnarray}
Y[A,\la]\equiv exp~\PLinv\Ga^{eff}[A,\la]\equiv
{}~exp~\PLinv\{~\Gahat[\la]+\la A~\} \nn\\
=\int\Dcal\vp~exp\ \frac{1}{\PL}[\ S_0[\vp]
-\la (\intx e^\vp - A)]   \com            \label{3.4k}
\end{eqnarray}
the stationary point $\la_c$\ of  $\Ga^{eff}[A,\la]$ is determined by
\begin{eqnarray}
\frac{d}{d\la}\Ga^{eff}[A,\la]|_{\la_c}
=\frac{d\Gahat[\la_c]}{d\la_c}+A=0\com\q
\la_c=\la_c^0+\PL\la_c^1+\cdots\pr\label{3.4l}
\end{eqnarray}
It gives the dominant contribution to the contour integral of (\ref{3.4cx}).
This condition (\ref{3.4l})
coincides with the equation (\ref{3.4c}) if we identify
$A$\ with $<A_{op}>$. It means the dominant contribution to the contour
integral comes from the value of $\la$\ at which the grand partition function
takes
$<A_{op}>=A$.
\footnote{
$\Ga^{eff}[A,\la_c]$\ is exactly
the same as the ordinary  (Schwinger's) effective action which is
obtained by Legendre transformation of $\Zhat[\la]$\ due to the change of
the independent variable from $\la$\ to $A=<A_{op}>$.
          }
Finally we obtain the approximate relations,
\begin{eqnarray}
Z[A]\approx \PLinv Y[A,\la_c]\com
\q Y[A,\la_c]=\ exp~\PLinv\Ga^{eff}[A,\la_c]
\approx exp~\PLinv \{ \Gahat^0[\la^0_c]+\la^0_c A\}\com    \label{3.4n}
\end{eqnarray}
where the former approximation
is valid in the large system limit and the latter one is
valid in the semi-classical limit.
In the following, we will evaluate the leading part (order of
$\hbar^0$) of $\Ga^{eff}[A,\la_c]$\ :\
$ \Gahat^0[\la^0_c]+\la^0_c A =\ S_{\la^0_c}+\la^0_cA\ $.
\section{Classical Configuration of R$^2$-Gravity and$\q$
         Phase Structure}
\subsection{Classical Solution}
The classical solution for
$\be=0$\  has been known as the Liouville solutions.
(See ref.\cite{SEI} for a recent review.) Furthermore,
in the context of 2d quantum gravity or the string theory ,
it was
studied by \cite{OV} and \cite{Z} .
We consider here the general case of $\be$\ being an arbitrary real number.
The classical equation
\ ,$\frac{\del S_\la[\vp]}{\del\vp}=0$\ ,
 is  explicitly written as
\begin{eqnarray}
\frac{\del S_\la[\vp]}{\del\vp}=\frac{1}{\ga}\pl^2\vp
+\be\{ e^{-\vp}(\pl^2\vp)^2-2\pl^2(e^{-\vp}\pl^2\vp)\}-\la e^\vp=0\pr
                                                         \label{3.6a}
\end{eqnarray}
We make the  assumption of constant curvature
for the  solution.
\footnote{
The importance of the constant-curvature solution will be commented on
in Sect.6. Other solutions will not be considered. They correspond to
different (classical) vacua from the present one.
}
\begin{eqnarray}
-R|_{\vp_c}=\
e^{-\vp_c}\pl^2\vp_c
=\mbox{const}\equiv \frac{-\al}{A}\com        \label{3.8}
\end{eqnarray}
where $\al$ is a dimensionless constant
which should satisfy
\begin{eqnarray}
\mbox{COND.1}\qqqq\al^2\be'-\frac{1}{\ga}\al-\la A=0\com\q
\be'\equiv\frac{\be}{A}\com
                                               \label{3.9a}
\end{eqnarray}
as the consequence of classical field equation (\ref{3.6a}).
It has  real solutions  $\al$\ when parameters
 $\be',\la$\ and $\ga$ \  satisfy
$D_1\equiv\frac{1}{\ga^2}+4\be'\la A \geq 0$\ .
Since
eq.(\ref{3.8}) is
the Liouville equation
with the
cosmological constant $-\frac{1}{\ga}\frac{\al}{A}$
(which is negative for $\al>0$ and positive for $\al<0$\
in the present case of $\ga>0$\ ),
the present solution contains that of Refs.\cite{OV,Z} as the $\be=0$
case.

\qq In this paper we consider only the case of
the positive curvature:\
$\al >0$\ .
The spherically symmetric
\footnote{
in the ($x^1,x^2$)-plane
}
 solution of (\ref{3.8}) is known to be
(cf.\cite{OV,Z,SEI}),
\begin{eqnarray}
\vp_c(r;\al )=-ln~\{ \frac{\al}{8}(1+\frac{r^2}{A})^2\}\com\q
r^2=(x^1)^2+(x^2)^2\pr  \label{3.10a}
\end{eqnarray}
It gives
$\left.\intx\sqg R\right|_{\vp_c}=-\intx~\pl^2\vp_c=8\pi$\ ,
which says the manifold described by the solution (\ref{3.10a})
has the sphere topology. The area,
$\left.\intx\sqg\right|_{\vp_c}=\intx~e^{\vp_c}=\frac{8\pi}{\al}A$\ ,
can be interpreted as the effective area covered by the classical solution.
The equations (\ref{3.9a}
-\ref{3.10a}) constitute a solution of (\ref{3.6a}).

\qq $S_{\la}[\vp_c]$ is given as
\begin{eqnarray}
& S_{\la}[\vp_c]
=(1+\xi)\frac{4\pi}{\ga}~ln\frac{\al}{8}-16\pi\al\be'
+C(A)\com &                                          \nn\\
& C(A)=\frac{8\pi (2+\xi)}{\ga}+\frac{8\pi\xi}{\ga}
\{~ln(1+L^2/A)-(L^2/A)/(1+(L^2/A))~\}\com      &
                                             \label{3.11}
\end{eqnarray}
where
$L$\ is the infrared cut-off
($r^2\leq L^2$)
introduced for the divergent volume intgral of the total derivative term.
Note that $C(A)$\ does not depend on $\be$ and $\al$\ .
For the analysis of the $\be$-dependence of physical quantities,
we may disregard $C(A)$. However ,for the $A$-dependence (such as that of
$Z[A]$\ ),
$C(A)$\ plays an important role.
The eq. (\ref{3.4l}) at the classical level is written as,
\begin{eqnarray}
\frac{dS_{\la}[\vp_c]}{d\la}+A
=\{ \frac{4\pi}{\ga}\frac{1}{\al}(1+\xi)-(16\pi\be'+\frac{1}{\ga})
+2\be'\al\}\frac{d\al}{d\la}=0\com                   \label{3.16}
\end{eqnarray}
where we have used a relation :\ $1=\frac{d\la}{d\al}\frac{d\al}{d\la}
=\frac{1}{A}(~2\al\be'-\frac{1}{\ga})\frac{d\al}{d\la}\ $
, which is derived
from (\ref{3.9a}).
This equation fixes the stationary point which dominates in the contour
integral (\ref{3.4cx});
\begin{eqnarray}
& \mbox{COND.2}\qqq 2\be'\al^2-(16\pi\be'+\frac{1}{\ga})\al+
(1+\xi)\frac{4\pi}{\ga}=0\com  &\label{3.17}
\end{eqnarray}
which has two real solutions;
\begin{eqnarray}
& \al^{\pm}_c=\frac{1}{4\be'}\{ 16\pi\be'+\frac{1}{\ga}
\pm\sqrt{(16\pi\be')^2+\frac{1}{\ga^2} -\xi\frac{32\pi}{\ga}\be'} ~\}
                                      \com       &  \label{3.17a}
\end{eqnarray}
when the condition
$ D\equiv (16\pi\be')^2
+\frac{1}{\ga^2}-\xi\frac{32\pi\be'}{\ga}
=(16\pi\be'-\frac{\xi}{\ga})^2+\frac{1-\xi^2}{\ga^2}\geq 0 $
is satisfied.
The relation (\ref{3.9a}) then determines $\la^{\pm}_c(\be)
\equiv \la(\be,\al^{\pm}_c(\be))$. Note that the determinant of the above
quadratic equation  is positive definite for all real
$\be$ if we take $\xi$\ for the region\ :\
$-1\leq\ \xi\ \leq\ +1$\ .
We consider this case in the following.

$\qq$ In summary two unknown parameters $\al$\ and $\la$\ are fixed by
two conditions COND.1 and 2 ,and they are expreesed by three physical
parameters
\ $\be$\ ,$\ga$\ ,$A$\ and one free parameter\ $\xi$.
In Fig.1 we plot $\al^\pm_c$\ ,which is equal to the curvature$\times A$\ ,
as the function of $w\equiv 16\pi\be'\ga$\ .
The solution of $\al^+_c$ is negative in the region of $\be<0$\ .
This contradicts the present condition $\al>0$\ .
Furthermore
the curvature and other physical quantities
,calculated using $\al^+_c$\ ,  diverge as $\be\rightarrow \pm 0$\ .
These behaviours contradict the results of numerical simulation.
Therefore we consider mainly $\al^-_c$-solution in the following.
( $\al^+_c$-solution will be discussed in sect.5, in relation to KN's result.)

{\vs 6}
\begin{center}
Fig.1\q $A\times$ Curvature
\ ,$\al^\pm_c$-branches,\ $w\equiv 16\pi\be'\ga,\ \xi=0$
\end{center}

\subsection{Analysis of $\al^-_c$-Solution and Cross-Over Phenomenon}

In Fig.2  the $R^2$-expectation value \ :\ $A<\intx \sqg R^2>$\  =
$-\frac{\pl \Ga^{eff}[A,\la_c]}{\pl\be'}$\  is shown in the Log-Log scale
for $w>0$\ .
It clearly shows the transition
similar to one observed in the numerical simulation.
Later (in Fig.5) we will show the theoretical curve in the linear scale
for all real $w$\ .
This classical
solution gives rather good agreement even in the negative $w$\ region.
Fig.3 and Fig.4 show the string tension $\times \ga A=\ \ga\la_c A$\ ,and
\ the total free energy $ \times \ga \
=-\ga \Ga^{eff}[A,\la_c]$\ ,respectively.
\footnote{
As for the figure of the total free energy (Fig.4), the $\be$-independent part
$C(A)$ is omitted.
}
{}From the fact that the effective area is given by the inverse of $\al^-_c$\
(\ the effective area $\times \frac{1}{A}\
=\frac{1}{A}\intx e^{\vp_c}\ ={8\pi}/{\al^-_c}$\ )
and from the behaviour of $\al^-_c$\ in Fig.1, we notice
that the area covered by the classical configuration is not
the same as as the area constrained by the $\del$-function in the micro-
canonical partition function except
the $\be\rightarrow -\infty$ region. This has happened because we are
approximating the fully-quantumly fluctuating manifold by
a simple classical sphere whose configuration is specified only by the
effective area
$1/\al$\ and the string tension $\la$\ .
This characteristically shows the present effective action approach using
$Y[A,\la]$\ (\ref{3.4k}). This point will be discussed further in sect.5.

{\vs 7}
\begin{center}
Fig.2\q $A<\intx\sqg R^2>|_c$,\ Log-Log plot for $w>0$
\ ,$\al^-_c$-branch,\ $\xi=0$
\end{center}
{\vs 6}
\nopagebreak[4]
\begin{center}
Fig.3\q$\ga A\times$(String Tension)
\ ,$\al^-_c$-branch,\ $\xi=0$
\end{center}
{\vs 6}
\begin{center}
Fig.4\q$\ga \times$(Total Free Energy)
\ ,$\al^-_c$-branch,\ $\xi=0$
\end{center}

$\qq$ The asymptotic behaviours of some  physical quantities
are listed in Table 1.

\vspace{0.5cm}
\begin{tabular}{|c|c|c|c|c|}
\hline
Phase      & (C) $w\ll -1$  & \multicolumn{2}{c|}{(B)  $|w|\ll 1$}
                                              & (A) $1\ll w$          \\
\hline
$\al^-_c$  & $8\pi$
                           & \multicolumn{2}{c|}{ $4\pi(1+\xi)\{1 $ }
                                             & $\frac{4\pi(1+\xi)}{w}$ \\
           & $+O(|w|^{-1})$ & \multicolumn{2}{c|}{$-\frac{1-\xi}{2}w\}+O(w^2)$}
                                                    & $+O(w^{-2})$     \\
\hline
$-\frac{\pl \Ga^{eff}_-}{\pl\be'}$
           & $64\pi^2+\frac{0}{|w|}$
                            & \multicolumn{2}{c|}{ $16\pi^2(1+\xi)\{3-\xi$  }
                                            & $\frac{64\pi^2(1+\xi)}{w} $ \\
           & $ +O(w^{-2})$
                            & \multicolumn{2}{c|}{ $-(1-\xi)^2w\}+O(w^2)$ }
                                                       & $+O(w^{-2})$   \\
\hline
$\ga\la^-_cA$ & $-4\pi |w|\{1$
                            & \multicolumn{2}{c|}{ $4\pi(1+\xi)\{-1$  }
                                          & $-\frac{\pi}{w}(1+\xi)(3-\xi)$ \\
          & $+O(\frac{1}{|w|})\}$
                          & \multicolumn{2}{c|}{ $+\frac{3-\xi}{4}w\}+O(w^2)$ }
                                                           & $+O(w^{-2}) $ \\
\hline
$-\ga\Ga^{eff}_-$
          & $-4\pi |w|\{1$ & \multicolumn{2}{c|}{
                                         $4\pi(1+\xi) \{1-ln~\frac{1+\xi}{2}$
                                                 }
                                                     & $4\pi(1+\xi)~ln~w$   \\
          & $+O(\frac{1}{|w|})\}-\ga C(A) $
            & \multicolumn{2}{c|}{ $+\frac{3-\xi}{4}w\}+O(w^2)-\ga C(A)$ }
                                           & $+$ const\ $-\ga C(A)$      \\
\hline
\multicolumn{5}{c}{\q}                                                 \\
\multicolumn{5}{c}{Table 1\ \  Asymp. behaviour of
                               physical quantites for $\al^-_c$-solution.}\\
\multicolumn{5}{c}{\q
   $R>0, w\equiv 16\pi\be'\ga, \ga=\frac{48\pi}{26-c_m}>0\ (c_m<26)$.
$C(A)$\ is given by (\ref{3.11}).                                         }\\
\end{tabular}
\vspace{0.5cm}

$\qq$ From these graphs and Table 1, we can observe three types of surfaces.
\flushleft{(A)\ Free Creased Surface;\ Large\ positive\ $\be\ (w\gg 1)$\ }

As $\be$\ increases,
the string tension decreases to zero (in the negative sign),
$\ga\la^-_c A\sim -\frac{\pi(1+\xi)(3-\xi)}{w}$,
and dynamics is mainly governed
by the 'kinetic' and total derivative terms (\ref{3.2}).
This phase is not influenced by the area condition or the 'potential' term
$-\la e^\vp$\ in (\ref{3.4k}) or (\ref{3.4g}).
The characteristic mass scale
is $1/\sqrt{\be}$\ as shown in the asymptotic
behaviour of Rieman curvature $R\propto\frac{1}{\be}$\ and of the string
tension $\la^-_c\propto -\frac{1}{\be}$\ .
The asymptotic behaviour
$Z[A]\sim A^{4\pi(1-\xi)/\ga}\times e^{O(1/w)}$\
shows the conformal behaviour .
The surface is mildly 'creased' with the curvature of order
$\frac{1}{\be}$ \ . As $\be$\ increases, the size of the creases
on the surface becomes large (the surface becomes less creased)
and the 'effective area' increases. As $\be$\ decreases, the surface
becomes more creased and the 'effective area' decreases.
$\frac{1}{A}\intx~e^{\vp_c}=\frac{8\pi}{\al}\sim 2w/(1+\xi)$\
shows this situation.
The data of the simulation well fits with the above image. Firstly the
predicted asymptotic behaviour
$A<\intx\sqg~R^2>\sim \frac{64\pi^2}{w}(1+\xi)$\  well describe the data
both qualitatively and quantitatively.
(We will soon do the fitting with data in Sect.4.)
Secondly the loop-length distribution\cite{TY}
and the coordination number distribution\cite{YTS}
clearly shows the above image.
\flushleft{(B)\ Fractal surface;\ $\be\approx 0\ (|w|\ll 1)$}

The string tension is finitely present and the sign is negative:
$\ga\la^-_cA\sim -4\pi(1+\xi)+3\pi(1+\xi)(3-\xi) w$\ .
The surface configuration is determined
not only by the 'kinetic'  and total-derivative terms
but also by the 'potential' term.
The two mass parameters (the coupling
$\be$\ and the area parameter $A$\ ) are balanced in such a way that
there is no charactersic mass-scale in this phase.
All physical quantites behave linearly with respect to w. In particular
the asymptotic behaviour:
$A<\intx\sqg~R^2>\sim 16\pi^2(1+\xi)\{ 3-\xi-(1-\xi)^2 w\} $\
well describes the data
of the computer simulation both qualitatively and quantitatively.
The behaviour $Z[A]|_{w=0}\sim A^{-8\pi\xi/\ga}$\ shows the conformal
one.
The value of the curvature at $\be=0$\ is
$R\times A|_{w=0}=\al^-_c(w=0)=4\pi(1+\xi) $\ .
\footnote{
This value is compared with
the expectation value obtained from the known
exact coordination-number($q_i$) distribution on lattice:\
$R_ia^2=2\pi\frac{6-q_i}{q_i},\ a^2=\mbox{unit area of a triangle},
<R_ia^2>=2\pi\sum_{q=3}^{\infty}
\frac{6-q}{q} W(q)\approx 4\pi\times 0.117478\com\
W(q)=16\cdot(\frac{3}{16})^q\cdot\frac{(q-2)(2q-2)!}{q!(q-1)!}\pr$
\cite{BK}
}
The cross-over point between (B) and (A) is
roughly obtained
as the point where  the approximation-condition for this region
breaks down:\ $w_{C.O.}=16\pi\ga\be'_{C.O.}\sim 1$.
(We will soon define the point definitely and obtain the explicit expression .)
Note that the cross-over point on $\be'$-axis goes to $+\infty$\ as
$c_m\rightarrow -\infty$\ (so-called 'classical'
limit in 2d quantum gravity):\
$\be'_{C.O}\sim\frac{1}{16\pi\ga}=\frac{26-c_m}{16\times 48\pi^2}\rightarrow
+\infty,\ c_m\rightarrow -\infty$\ .
\flushleft{(C)\ Strongly-Tensed Perfect Sphere;\
Large negative\ $\be\ (w\ll -1)$\ }

The constant value of the curvature $\al^-_c\sim 8\pi$\
,irrespective of the value $\xi$\ ,
implies this phase  describes the 'perfect sphere'.
\footnote{
This terminology 'perfect sphere' is used here in order to discrminate
the configuration that the surface is, as its shape, a sphere
from the configuration that the surface is topologically a sphere.
}
The asymptotic behaviours
$\la\propto -\frac{|\be|}{A^2}\ ,\ R\propto \frac{1}{A}\ $\ show
the characteristic mass scales are $\frac{1}{\sqrt{A}}$\ in addition to
$\be$. Dynamics is strongly influenced by the potential term.
Both the string tension and the total free energy are
negatively divergent as $\be\rightarrow -\infty$.
The surface is strongly tensed.

\section{Role of Total Derivative Term ,Determination of $\xi$ and Data Fit }
\qq Let us see more closely
how much the present analytical prediction fits with
the data and see the role of the total derivative term(
$\xi$-term in (\ref{3.4g})\ ).
All the graphs in sect.3 are evaluated at $\xi=0$.
The log-log plot of
$-\frac{\pl \Ga^{eff}[A,\la_c]}{\pl\be'}$ (Fig.2) shows, at some point
$w_c>0$, the behaviour qualitatively
changes from the linearly-descending line to
the constant-line as we decrease $w$.
We call the changing point ,$w_c$\ ,{\it cross-over point}.
Let us define the point definitely and
see its $\xi$-dependence.
Those two straight lines are given
as
$-\frac{\pl \Ga^{eff}[A,\la_c]}{\pl\be'}
\rightarrow 64\pi^2\frac{1+\xi}{w}\q\mbox{as}\q w\rightarrow +\infty
\ ,\
-\frac{\pl \Ga^{eff}[A,\la_c]}{\pl\be'}
\rightarrow 16\pi^2(1+\xi)(3-\xi)+O(w)\q\mbox{as}\q w\rightarrow +0
$\ .
We can unambiguously define
the crossing point  of two asymptotic lines above as
the cross-over point $w_c$\
, and get as
\ $w_c(\xi)=\frac{4}{3-\xi}
$\ .
$w_c$\ moves in the range $1\leq w_c \leq 2$\ for the present choice
of $\xi$\ :\
$-1\leq\ \xi\ \leq\ +1$\ .
This result shows the $\xi$-term determines the essential part of the theory.

\qq For the $\be=0\ (w=0)$\ case, the partition function is exactly known as
$Z_{exc}[A]=A^{\ga_s-3}, \ga_s=(c_m-1-\sqrt{(1-c_m)(25-c_m)}~)/12$.\cite{KPZ}
\ The present approximate result should coincide with it at the 'classical'
limit,$c_m\rightarrow -\infty$. This requirement gives $\xi=1$\ .

\qq Now we fit the present theoretical curve
of \ $A<\intx \sqg R^2>$\
with data of \cite{TY}.
Three adjusting
parameters ($P_1,P_2,P_3$) are necessary for the fit:
\begin{eqnarray}
&-\frac{\pl \Ga^{eff}[A,\la_c]}{\pl\be'}=\ P_1\cdot Y\com
\q w=\ P_2\cdot (X+P_3)\com\q     & \label{3.22}
\end{eqnarray}
where $(w,-\frac{\pl \Ga^{eff}[A,\la_c]}{\pl\be'})$\ is the theoretical scale
(see Table 1)
and $(X,Y)$\ is the scale of the simulation data. The meaning of the
adjusting parameters are as follows:\
1)\ $P_1$\ adjusts the scale of
the expectation value ,
$<\intx\sqg R^2>$,\ itself
\footnote{
$P_1$\ is the ambiguity of a multiplicative constant which appears in
comparing an expectation value of a continuous theory with that of the
corresponding lattice theory.
}
\ ;\
2)\ $P_2$\ adjusts the scale of the 'width' of the phase (B) in the w-axis\ ;\
3)\ $P_3$\ adjusts the origin of the w-axis.
\footnote{
This parameter $P_3$\ reflects the renormalization (quantum) effect.
}
$P_1$\ and $P_2$\ should be positive, whereas $P_3$\ may be positive,zero or
negative.
We can fix those  parameters ,for each $\xi$\ ,by the use of
three data points:\
$(X,Y)=(-100.0,1.69265),(0.0,0.70605),(100.0,0.08781)$.
In Fig.5, we plot the adjusted curves of
$-\frac{\pl \Ga^{eff}[A,\la_c]}{\pl\be'}$ \
,in the linear scale,
for three typical values of $\xi$\
$(-0.99,0.0,0.99)$\ with the simulation data. The parameters used
in Fig.5 are listed in Table 2.
We must realize that the  the total derivative
term  greatly influences the final result.
{\vs 7}
\begin{center}
Fig.5\ \ Fit of $<\intx\sqg R^2>$\ .Dots are data points.
\end{center}

\vspace{0.5cm}
\begin{tabular}{|c|c|c|c|}
\hline
$\xi$  & (1)\ -0.99 & (2)\ 0.00  & (3)\ 0.99            \\
\hline
$P_1$ &  373.2    & 373.2    &   373.2               \\
\hline
$P_2$ &  4.505$\times 10^{-3}$  & 0.1705 &  0.3365   \\
\hline
$P_3$ & -284.6   & 8.556 & 12.48    \\
\hline
\multicolumn{4}{c}{\q}                                   \\
\multicolumn{4}{c}{Table 2\ \  Parameters used in  the data fit of Fig.5
                                                           }\\
\end{tabular}
\vspace{0.5cm}

\section{$\al^+_c$-Solution and Kawai-Nakayama's Result}
\qq
There exits a conformal approach to the present problem \cite{KN}.
They treat the phase (A) in Table 1 as the conformal phase.
Their result about the asymptotic behaviour of the partition function $Z[A]$\
does not coincide with the present one.
We discuss the origin of the discrepancy. The sharp contrast of the two
approaches exists in the treatment of the area constraint:\
$\intx\sqg =A$\ , and the topological constraint:\
$\intx\sqg R=8\pi$\ . (i) The present approach does not directly 'solve'
the area costraint, whereas KN does it. (ii) We respect the topological
constraint, whereas KN does not.

$\q$
For (i), we introduce the parameter of the chemical potential $\la$\ ,which
can be regarded as the 'Lagrangian multiplier' for the area-constraint
as shown in (\ref{3.4k}) and is physically interpreted as the surface (or
string) tension.
The validity of this treatment in the semiclassical approach can be stated as
follows. The 'effective' sphere (the classical solution (\ref{3.10a})),
which approximates the fully-quantum
surface-configuration ,does not necessarily
satisfy the area constraint $\intx\sqg=A$\ . The constraint is satisfied
only when the  dominant configuration is near the perfect sphere
which characteristically has the large surface
-tension (positive or negative) and  the characteristic mass scale of
$\frac{1}{\sqrt{A}}$.
\footnote{
Phase (C) in sect.3 is the case. In the phase (B), the surface-configuration
is {\it not} near the perfect sphere. In this phase, however, the area
constraint is satisfied by virtue of the 'topological effect'
due to $\xi$-term\ :\
$\intx \sqg|_{\vp_c}\approx \frac{2A}{1+\xi}=A$\ for $\xi=1$.
}
When the surface-tension is not large,
the configuration is far from the perfect sphere and
we cannot use the area-constraint on the leading configuration. In other words,
the area-cnstraint must also be treated  'perturbatively' as far as
the semiclassical approximation works correctly. For (ii), we have introduced
the parameter $\al$\ in (\ref{3.8}) to give the variableness for the value
of the constant curvature. This variableness gives, through the solution
(\ref{3.10a}), the correct constraint for the topological quantity:\
$\intx\sqg R|_{\vp_c}= R|_{\vp_c}\cdot\intx\sqg|_{\vp_c}=
\frac{\al}{A}\times\frac{A}{\al}8\pi=8\pi\ $.

$\q$
In the  analysis  of previous sections,we have considered only $\al^-_c$-
solution which does not satisfy the area constraint for $w\gg 1$\ (A-phase)\ :
\ $\intx\sqg|_{\vp_c}\approx A\times \frac{2w}{1+\xi}$\ . As for $\al^+_c$-
solution, the following asymptotic behaviours are obtained for $w\gg 1$.
\begin{eqnarray}
& \al^+_c=8\pi+O(w^{-1})\  ,\
\intx e^{\vp^+_c}=A(1+O(w^{-1}))\ ,\
\ga \la^+_c A=4\pi w(1+O(w^{-1})), & \nn\\
&\mbox{for}\qq w\gg 1\pr & \label{5.1}
\end{eqnarray}
This result shows, $\al^+_c$-solution satisfies the
area constraint for $w\gg 1$.
We explain below that this phase describes the perfect sphere.
As expected, we find exactly
KN's result in this region.
\footnote{
The relation between the present notation and the KN's is\
$32\pi\be=1/m^2$,where $1/m^2$\ is the KN's notation for the higher
derivative couplings.
}
\begin{eqnarray}
& Z[A]\approx A^{-\frac{8\pi \xi}{\ga}}exp\{-\frac{4\pi w}{\ga}(1+O(w^{-1}))\}
\com\q \xi=1                                             \pr \label{5.2}
\end{eqnarray}

$\q$ We make remarks about other properties of the branch $\al^+_c$.
The asymptotic behaviours are listed in Table 3.

\vspace{0.5cm}
\begin{tabular}{|c|c|c|c|c|}
\hline
 Phase & $w\ll -1$ & $-1\ll w<0$ & (E) $0<w\ll 1$ &  (D) $1\ll w$        \\
\hline
$\al^+_c$ &$<0$\ ,
     & $<0\ ,$
         & $\frac{4\pi}{w}\{2+w(1-\xi)$
            & $8\pi $    \\
 &\ not allowed & \ not allowed
         &  $+O(w^2)\}$ & $+O(w^{-1})$    \\
\hline
 $-\frac{\pl \Ga^{eff}_+}{\pl\be'}$ &/
     & / & $-\frac{64\pi^2}{w^2}\{ 1-(1+\xi)w$
                       & $64\pi^2\{1+\frac{0}{w}$          \\
  &  & &$\ \ +O(w^2)\}$  & $+O(w^{-2})\}$              \\
\hline
 $\ga\la^+_cA$&/
      & / & $\frac{4\pi}{w}\{-1+0\cdot w$
             & $4\pi w\{1$     \\
 & & & $+O(w^2)\}$ & $+O(w^{-1})\}$       \\
\hline
 & /&/& $\frac{4\pi}{w}\{1+2w$ & $4\pi w\{1 $      \\
 $-\ga \Ga^{eff}_+$ &  & & $+(1+\xi)w~ln~w$ & $+O(w^{-1}\} $  \\
 & & & $+O(w^2)\}-\ga C(A)$ & $-\ga C(A) $          \\
\hline
\multicolumn{5}{c}{\q}                                   \\
\multicolumn{5}{c}{Table 3\ \  Asymp. behaviour of physical quantities
                  for $\al^+_c$-solution.}\\
\multicolumn{5}{c}{
$R>0, w\equiv 16\pi\be'\ga, \ga=\frac{48\pi}{26-c_m}>0\ (c_m<26)$.
$C(A)$ is given by (\ref{3.11}).}
\end{tabular}
\vspace{1cm}

Each phase in Table 3 is explained as follows.

\flushleft{(D)\
Explosive Perfect Sphere;\
Large\ positive\ $\be\ (w\gg 1)$  }

This phase describes the configuration of the
strongly-expanding perfect sphere.
The asymptotic behaviour of the string tension:\
$\ga\la^+_c A\sim 4\pi x\rightarrow +\infty (x\rightarrow +\infty)$, shows
the surface is strongly forced expansively.
\footnote{
The phase (C) in sect.3 also describes the configuration of perfect sphere,
but the string tension and the total free energy have the different sign.
}
The asymptotic behaviour of the partition function is given above ,(\ref{5.2}).
The constant value of the curvature
($\al^+_c\sim 8\pi$) corresponds to the perfect sphere
with the radius$\sim\sqrt{A}$\ .
The characteristic length scale is fixed by the area parameter $A$
,not by $\be$.
The total free energy is
positively divergent ($-\ga \Ga^{eff}_+\sim 4\pi w$) as $\be$\
increases to $+\infty$, therefore this configuration is not preferable.
The predicted result,
$A<\intx\sqg~R^2>\sim (64\pi^2)$\ (constant), contradicts
the data of the lattice simulation\cite{TY}.
We conclude this phase does not describe the data.
\flushleft{(E)\ Degenerate Surface;\
Small\ positive\ $\be\ (0<w\ll 1)$}

As $\be$\ goes to $+0$,
the curvature increases to $+\infty$\ ($\al^+_c\sim \frac{8\pi}{w}$)
and the area decreases to $+0$\ ($\frac{8\pi}{\al^+_c}\sim w$).
This shows the surface is degenerate.
\footnote{
The behaviour of vanishing area makes us imagine that this phase
describes,so-called, branched polymer.
}
The radius of the 'effective' sphere is approximately $\sqrt{\be}$\ .
The characteristic length scale is controlled by $\sqrt{\be}$,
not by $\sqrt{A}$.
The string tension becomes negatively divergent ($\ga\la^+_c A\sim
-\frac{4\pi}{w}$). It means the (degenerate) surface is strongly tensed.
The partition function behaves asymptotically as
$Z[A]\sim A^{-\frac{8\pi\xi}{\ga}}exp(-\frac{4\pi}{\ga}\frac{1}{w})$,
which contradicts  the known conformal result.
The total free energy becomes positively divergent as $\be\rightarrow +0$:\
$-\ga\Ga^{eff}_+\sim +\frac{4\pi}{w}\rightarrow +\infty$\
, therefore this phase is
energetically unpreferable.
The predicted behaviour of
$A<\intx\sqg~R^2>\sim -\frac{64\pi^2}{w^2}$\ contradicts the simulation data .
It is crucial that the solution is not connected with the
$\be<0$\ region, while the simulation data  shows the physical quantities
are continuously
connected with the $\be<0$\ region.
This phase does not describe the simulation
data.

\vspace{0.7cm}
$\q$ As explained in (E), $\al^+_c$-solution does  have the bad behaviour for
$w\ra +0$\ ,which cannot be accepted in the conformal approach. Similar
bad behaviour is also noticed in KN's case.


$\q$ $\al^+_c$-solution shares similar properties with those of KN's.
KN's solution looks to correspond to the $\al^+_c$-solution in the
present formalism, in particular for the $w\gg 1$\ region.

\vspace{0.5cm}

\section{Discussions}
\qq Some additional comments are in order.

\begin{enumerate}
\item
We have pointed out the importance of the total derivative term in (\ref{3.2}).
There could be many other types of total derivative terms, but they are
exculded as follows.
\begin{enumerate}
\item
Higher-Derivative Terms:$\qq$
{}From the dimensional analysis,
the higher-derivative terms vanish for the limit $L\ra +\infty$
,where $L$\ is the infrared-regularization parameter introduced
in (\ref{3.11}).
For example:\
$\intx \pl_a(\vp_c\pl^2\pl_a\vp_c)\sim\ (ln~L)/L^2,\
 \intx \pl_a(\pl^2\vp_c\cdot\pl_a\vp_c)\sim\ 1/L^2$.
\item
Terms of Higher-Power of $\vp$\ :$\qq$
We may impose, on the acceptable 'topological' action of
$\intx \pl_a(\pl_a\vp\cdot \vp^n)$\ ,the natural condition that the critical
behaviour should not be influenced by the change of the regularization
parameter:\ $L\ra \mbox{const}\times L$\ . This condition uniquely fix the
power as $n=1$.\
(\ $\intx \pl_a(\pl_a\vp_c\cdot{\vp_c}^n)\sim\ L\cdot (1/L)(ln~L/A)^n
,L\ra +\infty$.\ )\
Note that $\intx\pl^2\vp_c=-\intx\sqrt{g} R|_{\vp_c}=-8\pi$.
\end{enumerate}
Therefore no ambiguity exists, except the $\xi$-term, in the theory.
\item
In the analysis of the ordinary conformal approach
,the kinetic term of $\vp$\ in
Liouville action is used  only for the explanation of assigning the free-field
form to the 2-point function of $\vp(x)$:\
$<\vp(x)\vp(y)>\sim\ ln~|x-y|$.\ The global(topological) effect,which is
essential for the critical exponents such as the string susceptibility,
is obtained not by the lagrangian but by the requirement of the conformal
symmetry (for the partition function).
The present approach contrasts with this.
We do not use the requirement of the conformal symmetry.
Instead we directly use the lagrangian and its explicit
Liouville solution which contains the essential part of the conformal symmetry.
And the global aspect of the theory can be taken into account through the total
derivative term in the lagrangian.
Note that the infra-red regularization $L$\ in (\ref{3.11}), besides the total
derivative term itself, is important for the quantity $Z[A]$\ to have
the correct conformal behaviour for $\beta$\ (or $w$) $\ra +0$.
The present analysis manifestly reveals
the importance of the infra-red regularization. This point was stressed by
T.Yoneya\cite{Y}.
\item
We have examined only the classical configuration in the present paper.
The quantum aspect is, of course, very important. One of us (S.I.) is preparing
for the quantum analysis in the present formalism\cite{SI2}.
$R^2$-term suppresses the ultra-violet divergences quite well and makes
the theory renormalizable. Besides the renormalizability, the unitarity
problem is also important generally in the higher-derivative theories.
In the present case of 2d $R^2$-gravity, it was argued in \cite{Y}
, for the case of Loretzian metric, that the framework for the unitarity
discussion ( such as the meaning of state, wave functinal, etc. ) should
be first settled. This problem deserves further study.
\item
We have chosen a constant curvature solution as the vacuum.
The importance of the constant curvature configuration in 2d quantum gravity
was stressed by A.H. Chamseddine\cite{CHA} in the context of the conformal
formalism of the ordinary (not $R^2$) 2d gravity. His model has
an auxiliary scalar field $\p$\ :\ $\Lcal_1=\sqg \p (R+\La)$\ .
Due to the presence of the auxiliary field,
the model always has, at the classical level,
a constant curvature configuration. He argues some difficulties in the
conformal approach, such as the limitation on the target-space dimension
in the string-terminology, are naturally resolved. $R^2$-gravity can be
regarded as a kind of the above-mentioned model\ :\
$\Lcal_2=\sqg \{ \p(R+\La)+c_1\p^2\}$. Kawai-Nakayama\cite{KN}
has taken this approach in their analysis.
Further generalization of the above model, which includes more-higher
derivative terms, has been studied in \cite{LO,MO,V,B}.
These general models are interesting as future alternate theories when
the 2d quantum gravity faces  serious problems.
\end{enumerate}

\qq We have analysed Liouville theory induced by $R^2$-gravity, at the
classical level.
For the analysis we have presented the effective action formalism
using $Y[A,\la]$\ (\ref{3.4k}),which efficiently takes into account
the area constraint.
The features of three phases are explained theoretically.
The importance of the total derivative term is stressed.
The free parameter $\xi$\ is fixed to be $1$\ by comparing the present
approximate result $Z[A]$\ with the exact KPZ result at the 'classical'
limit $c_m\rightarrow -\infty$\ .
In particular the prediction about the expectation value of
$<\intx\sqg R^2>$\ well fits the data of the computer simulation
for all real $\be$-region. It makes
sure of the validity of the semiclassical approach. The small discrepancy
comes from the quantum effect.

\begin{flushleft}
{\bf Acknowledgement}
\end{flushleft}
The authors thank Prof. T. Yoneya ,
Prof. H. Kawai , Prof. N. Ishibashi and Dr. A. Fujitsu
for discussions and comments .

\end{document}